\documentclass{elsart}  
  
\usepackage{natbib}

\usepackage{graphicx}

\usepackage{amssymb}  

\usepackage{color}  
  
\begin{document}  
  
\begin{frontmatter}  
  
  
  
\title{Nuclear break-up of $^{11}$Be}  
  
  
\author[ipno]{V.Lima},   
\author[ipno]{J.A.Scarpaci\corauthref{cor}},  
\corauth[cor]{Corresponding author.}  
\ead{scarpaci@ipno.in2p3.fr}   
\author[lpcc,ganil]{D.Lacroix},   
\author[ipno]{Y.Blumenfeld},   
\author[ipno]{C.Bourgeois},   
\author[ipno]{M.Chabot},  
\author[ganil]{Ph.Chomaz},   
\author[ipno]{P.D\'esesquelles},   
\author[ganil]{V.Duflot},   
\author[ipno]{J.Duprat},   
\author[ipno]{M.Fallot},   
\author[ipno]{N.Frascaria},   
\author[lpcc]{S.Gr\'evy},   
\author[ipno]{D.Guillemaud-Mueller},   
\author[ganil]{P.Roussel-Chomaz},   
\author[ganil]{H.Savajols},   
\author[ipno]{O.Sorlin}  
  
\address[ipno]{Institut de Physique Nucl\'eaire, Universit\'e Paris-Sud-11-CNRS/IN2P3, 91406 Orsay, France}  
\address[lpcc]{Laboratoire de Physique Corpusculaire, ENSICAEN and Universit\'e de Caen, IN2P3-CNRS, 14050 Caen Cedex, France}  
\address[ganil]{G.A.N.I.L., CEA et IN2P3-CNRS, BP 5027, 14076 Caen Cedex, France}  
\begin{abstract}  
% Text of abstract  
The break-up of $^{11}$Be was studied at 41 AMeV using a secondary beam of $^{11}$Be from the GANIL facility on a   
$^{48}$Ti target by measuring correlations between the $^{10}$Be core, the emitted neutrons and gamma rays.   
The nuclear break-up leading to the emission of a neutron at large angle in the laboratory frame is identified   
with the "towing mode" through its characteristic n-fragment correlation.  
%\textcolor{blue}{   
The experimental spectra are compared with a model where the time dependent Schr\"odinger equation (TDSE) is solved   
for the neutron initially in the $^{11}$Be.   
%to extract spectroscopic information on the wave-function in $^{11}$Be 
A good agreement   
is found between experiment and theory for the shapes of neutron experimental energies and angular distributions.  
The spectroscopic factor of the $2s$ orbital is tentatively extracted to be $0.46 \pm 0.15$. 
The neutron emission from the $1p$ and $1d$ orbitals is also studied. 
%A spectroscopic factor of  is found.  
%This led to the extraction of spectroscopic   
%factors for the GS of $^{11}$Be of 47\% of a $2s$ state coupled to a cold $^{10}$Be core and 50\% of a 1d state   
%coupled to an excited core. {\it VOIR SI ON MET ENTRE x\% et y\% ou SI ON MET DES BARRES D'ERREURS}.    
%}    
%The comparison of the neutron energy   
%spectra with a time dependent Schr\"odinger equation calculation led to the extraction of spectroscopic   
%factors for the GS of $^{11}$Be of 47\% of a $2s$ state coupled to a cold $^{10}$Be core and 50\% of a 1d state   
%coupled to an excited core.  
\end{abstract}  
  
  
\end{frontmatter}  
  
% main text  
\section{Introduction}  
\label{intro}  
The investigation of exotic nuclei has strongly developed during the last two decades thanks to the   
availability of radioactive beams produced by in-flight fragmentation \cite{tan85,han87,tan96} and   
structural properties of a large variety of nuclei far from stability can be accessed. Recent   
advances have demonstrated the importance of a detailed understanding of nuclear reactions in   
order to infer properties of nuclei such as nuclear shapes, spectroscopic factors or pairing   
correlations \cite{mar02}. Several reaction mechanisms are used : transfer reactions of one or two   
nucleons \cite{for99,dao04}, Coulomb or nuclear break-up reactions 
\cite{mar02,pal03,han03,sau04}, the latter being the focus of this paper.   
  
With exotic nuclei, we have to deal with low intensity beams and we expect an increasing   
complexity in obtaining nuclear structure information due to the reduction of the binding energy.   
In addition, modification of two-body correlations like pairing is expected.   
  
In this paper, we will demonstrate that the study of nuclear break-up around the Fermi energy,   
conjointly with the development of accurate nuclear reaction models \cite{lac99,esb01,kid94,typ01},   
can provide an alternative path to get information on structural properties in stable and weakly  
 bound exotic nuclei. In this respect, $^{11}$Be appears as a good case for study as it is a one-neutron   
halo nucleus and is abundantly produced. Its ground state is known to be a mixture of a cold $^{10}$Be   
core coupled to a halo neutron in a $2s_{1/2}$ state with an excited $^{10}$Be in a 2$^+$ state coupled to a   
1d$_{5/2}$ neutron leading to a 1/2 ground state with a positive parity.  
\begin{center}   
$|GS>$ = $\alpha$$|$0$^+$$\otimes$2s$_{1/2}$$>$ + $\beta$$|$2$^+$$\otimes$1d$_{5/2}$$>$  
\end{center}  
The determination of the relative importance of these two states   
%is an open question that   
was addressed through several experimental and theoretical approaches and is related to the   
spectroscopic factors S$_{2s}$=$\alpha$$^2$ and S$_{1d}$=$\beta$$^2$.  
$^{11}$Be was extensively studied during the last decade as one of the first observed halo nuclei   
\cite{tan85}. It has revealed many of its properties through several different types of reactions   
such as break-up experiments \cite{pal03,ann94,aum00,fuk04} or transfer reactions \cite{for99}. Coulomb   
dissociation has been shown to play an important role when $^{11}$Be interacts with a heavy target   
such as Au, as the halo neutron is bound by only 500 keV and easily separates from the core.   
  
For lighter targets such as Be, the nuclear break-up is expected to take over \cite{fal02}. It has   
been shown in a time dependent non-perturbative calculation (TDSE) \cite{lac99,esb01} that nuclear   
break-up presents a typical emission pattern, where the neutron and the remnant $^{10}$Be core are   
expected to be emitted in opposite sides of the beam direction. This property has already been   
observed with stable nuclei and was called "towing mode" \cite{jas98}. The pattern of this mechanism,   
leading to the emission of the particle at large angle, is expected to be sensitive to the   
initial wave function such as its angular momentum and could be used to shed some light on the   
structure of the nucleus.   
  
In this paper, we will report on an experiment performed at the GANIL facility on the break-up   
of $^{11}$Be where a $^{10}$Be fragment was detected in coincidence with a neutron and gamma rays. After   
a description of the experimental apparatus in section \ref{chap_setup} and of the theoretical framework in   
section \ref{chap_calc}, we will present the neutron angular distribution in paragraph \ref{chap-ds-dW} and 
demonstrate   
the characteristic anti-correlation between the neutron and the $^{10}$Be in paragraph \ref{chap-ds-dW-dte-gau}. 
The gamma   
energy distribution will be presented in paragraph \ref{chap-gamma}. Paragraphs \ref{en-2f} and \ref{en-3f} will be   
devoted to the neutron energy spectra in two-fold and three-fold events respectively. By   
comparing the experimental data with the time dependent Schr\"odinger calculation described   
in section \ref{chap_calc}, the possible extraction   
of spectroscopic factors of the ground state of $^{11}$Be will be discussed in section  
\ref{spefact}.    
  
\section{Experimental set-up}  
\label{chap_setup}  
  
\begin{figure}  
\begin{center}  
\includegraphics[width=11cm]{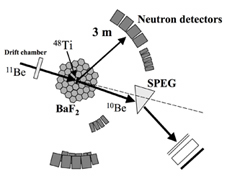}  
\end{center}  
\caption{Experimental set-up : The 74 gamma rays detectors were placed below and above the   
horizontal plane (37 detectors are shown on the plot). 23 neutron detectors were positioned   
around the target and the $^{10}$Be fragments were detected after analysis by the spectrometer SPEG.}  
\label{setup}  
\end{figure}

The reaction $^{48}$Ti($^{11}$Be,$^{10}$Be+n) at 41 MeV per nucleon was studied.   
The $^{11}$Be   
beam was produced by fragmentation of a 10 $\mu$A $^{13}$C$^{6+}$ primary beam onto a   
6128 $\mu$m thick carbon target placed in the Superconducting Intense Source for Secondary Ions   
device (SISSI) of the GANIL facility. A 600 $\mu$m thick aluminum degrader was set in the alpha   
spectrometer and a pure $^{11}$Be beam was transported to the SPEG vault with a small angular   
divergence of 0.2$^o$. The resulting beam intensity of 80000 pps was monitored by a drift   
chamber placed before the target. The beam size on the target was $\pm$6.4 mm giving an emittance   
after the break-up of more than 400 $\pi$.mm.mrad and a loss in detection efficiency of the SPEG   
\cite{bia89} spectrometer, which has a 70 $\pi$.mm.mrad nominal acceptance, of about a factor of 5.   
However, with the small angular divergence, no beam   
tracking was needed to extract the scattering angle of $^{10}$Be and hence to perform angular   
correlations between the ejectile and the neutrons. After breaking up on a 113 mg/cm$^2$ Ti target,   
the $^{10}$Be core was analyzed by the energy loss spectrometer SPEG positioned at 0 degree and covering $\pm$2$^o$   
in both horizontal and vertical directions. 74 BaF2 scintillators covered 84\% of the total solid   
angle around the target to detect the gamma rays and 23 neutron detectors were utilized and  
positioned as shown in the experimental set-up presented  
in Fig.\ref{setup}.  
  
\begin{figure}  
\begin{center}  
\includegraphics[width=8cm]{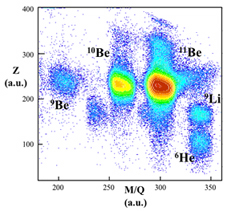}  
\end{center}  
\caption{Identification of the fragments produced in the Ti target and after the SPEG spectrometer.}  
\label{m-q-z}  
\end{figure}

Two drift chambers, an ionization chamber and a plastic detector forming the standard detection   
in the focal plane of SPEG were used for identifying the fragments and infer their momenta   
and scattering angles. An identification plot of the fragments produced after the Ti target is   
constructed by plotting the energy loss in the ionization chamber versus the arrival time of the   
ion in the plastic detector (proportional to the M/Q of the particle since the B$\rho$ is fixed by the   
spectrometer) and shown on Fig.\ref{m-q-z}. The spectrometer field was set to obtain the $^{10}$Be in the middle   
of the focal plane.   
  
\begin{figure}  
\begin{center}  
\includegraphics[width=8cm]{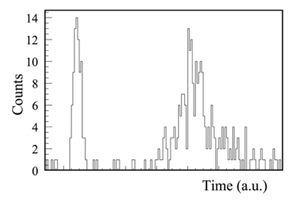}  
\end{center}  
\caption{Typical time of flight spectrum for the neutron detectors.}  
\label{Tn}  
\end{figure}

The neutrons were detected by 11 detectors of the NordBall array \cite{arn91} plus 8 cylindrical   
detectors of either 2 or 6 inches thick from the Uppsala group and 4, 30 cm long cylindrical detectors from   
Orsay. These neutron detectors of BC501 type were placed around the target at a distance   
varying from 1.4 meters to 3.5 meters (see Fig.\ref{setup}) and at angles running from 30$^o$   
to 95$^o$ in the laboratory frame. The identification was given by pulse shape   
discrimination by integrating the signal in two 400 ns wide gates, the first starting   
at the beginning of the signal and the second shifted by 30 ns. Energies were obtained   
from the time of flight (ToF) measured between the RF of the beam and the neutron arrival   
time in the detector. A typical ToF spectrum is presented in Fig.\ref{Tn}.   
  
74 BaF2 detectors of the Ch\^ateau de Cristal ensemble were used to detect the gamma rays.   
Detector efficiency and energy calibrations where performed with $^{60}$Co and $^{137}$Cs  
radioactive sources. GEANT3 was used to simulate the full detector and gave an   
overestimation of the experimental efficiencies obtained at 661 keV ($^{137}$Cs), 1173 and 1332   
keV ($^{60}$Co) by 23\%, as already observed in ref.\cite{mar00}. The detection efficiency of   
gamma rays of higher energies was then obtained by the GEANT3 simulation code (release 3.21).   
Furthermore, as the cells were generally not directed toward the target and were overlapping   
in angle, GEANT3 was also used to infer the effective angle at which each detector measured  
 the gamma rays coming from the target position. These angles, that were different from the average   
geometrical angle of each detector, were necessary for Doppler correction of the energy of the   
gamma rays emitted by the ejectile. As will be shown in paragraph \ref{chap-gamma}, three peaks   
corresponding to gamma rays from $^{10}$Be were clearly observed. Whenever these peaks could   
be seen for a single detector in coincidence with a $^{10}$Be, their positions were used for   
the energy calibration at high gamma ray energies. The energy range of interest in our   
case is about 3.37 MeV in the $^{10}$Be frame as it corresponds to the decay of the 2$^+$   
state to the GS (see paragraph \ref{chap-gamma}) which corresponds to 4.5 MeV in   
the laboratory frame at very forward angles.  
  
\section{TDSE Calculations}  
\label{chap_calc}  
  
During the past decade several approaches have been developed to describe   
the Coulomb and nuclear breakup at and above the Fermi energy.   
Among them, the Eikonal approximation, which takes advantage of the high velocity of   
the projectile, has encountered a large success in particular to interpret   
recent knock-out reactions (see for instance the recent review \cite{han03}). It is however worth mentioning   
that the "towing-mode" which is part of the diffractive dissociation channel mechanism is very   
specific. Indeed, as noted in ref. \cite{esb01}, description and understanding of the   
"towing mode" goes beyond calculations within the Eikonal approximation.   
On opposite, models based on the resolution of TDSE equations \cite{lac99,esb01,typ01,kid94,esb95,mel99,typ99, 
mel01,cap04}  
properly account for such a continuum emission. The great advantage of TDSE techniques is   
to treat non-perturbatively the nucleon emission to the continuum and the interferences between Coulomb and  
nuclear break-up.  
On the other hand, TDSE method often neglect interferences due to the mixing of different impact  
parameters that are generally treated properly in the Eikonal model.  
Only recently, methods have been developed that combines the advantages of both Eikonal and   
TDSE framework \cite{gol06}.

Here, we restrict ourselves to the simplified situation where  
a time dependent Schr\"odinger equation (TDSE) has been solved for a one-nucleon wave   
function initially in a $^{10}$Be core and a comparison with the experimental data will be   
presented in section \ref{spefact} to extract spectroscopic factors. In the present section we will briefly   
describe the calculation and more details can be obtained from ref.\cite{lac99, fal02}.   
  
The initial wave function of the halo neutron was deduced from the solution of the   
single-particle Hamiltonian that writes   
  
\begin{center}  
\begin{equation}  
h=\frac{\hat{p}^2}{2m}+V_{Be}(\hat{r})  
\end{equation}  
\end{center}  
  
where the $^{10}$Be potential was taken as a Wood-Saxon   
  
\begin{center}  
\begin{equation}  
V_{Be}(\hat{r})=\frac{V_0}{1+e^{\frac{(\hat{r}-R_{Be})}{a_0}}}  
\end{equation}  
\end{center}  
  
where the diffuseness $a_0$ = 0.75 fm and the radius $R_{Be}$ = 1.27.11$^{1/3}$ =2.82 fm. Numerically  
the initial wave-packet is obtained by diagonalizing the one-body Hamiltonian in spherical   
coordinates in a sphere of 30 fm radius with a space-step of $\Delta$r = 0.02 fm. The depth $V_0$   
was taken to be -46.025 MeV to finally obtain a $2s$ wave function bound by 0.503 MeV.  
  
The wave function calculated in spherical coordinates for $^{11}$Be is then transformed in   
Cartesian coordinates for the dynamical evolution and an appropriate interpolation is   
made. Special attention has been given to the purity of the ground state $^{11}$Be neutron   
halo by performing an additional imaginary time evolution on the Cartesian network, after   
which the $2s$ state was bound by -0.503 MeV.   
  
The dynamical evolution of the system is given by the following single particle   
Schr\"odinger equation;  
  
\begin{center}  
\begin{equation}  
i\hbar \frac{d}{dt}|\phi \rangle= \Big\{ \frac{\hat{p}^2}{2m}+V_T(\hat{r}-r_T(t))+  
V_{Be}(\hat{r}-r_{Be}(t)) \Big\} |\phi \rangle  
\end{equation}  
\end{center}  
  
where $V_T$ and $V_{Be}$ are the target and the projectile time-dependent potentials. The positions   
$r_T(t)$ and $r_{Be}(t)$ correspond to the target and the projectile respectively.  
  
The nuclear potential of the Ti target is taken to be a Wood-Saxon potential $V_T(r)$ with a   
diffuseness $a$ equal to 0.75 fm, a radius $R=1.17 A^{1/3}$ fm (A being the mass number of Ti) and a   
depth $V_0$ of -41.2 MeV adjusted to obtain the experimental binding energy of the last neutron   
in a 1f state which is -11.6 MeV. 
The sensitivity to different parameters  
of the calculation will be discussed in the following.  
  
In order to take into account the $^{10}$Be core displacement, the evolution is performed for   
each time step using the classical Coulomb trajectory for the center of mass motions from   
a distance of -400 fm along the initial velocity axis between the projectile and the   
target and equal to the impact parameter $b$ on the perpendicular axis. The Runge Kutta  
 numerical method is applied for the trajectory calculation. The split operator method \cite{fei82}   
is used to solve the time dependent Schr\"odinger equation with a time step of 1.7 fm/c on  
a mesh of 64x64x64 (fm$^3$) with a step size of 0.5 fm. This method precludes the inclusion of   
a spin-orbit term.

The calculation is then performed for several impact parameters with either a sharp cut-off or a weight  
that corresponds to a survival   
factor mimicking a probability that the core remains intact during the collision   
\cite{bon99}. This probability writes:  
 
%\begin{center}  
\begin{eqnarray}  
P(b)=exp\left({-ln 2 \times exp \left(\frac{ r_s-b}{a_s} \right)} \right) 
\end{eqnarray}  
%\end{center}  
where   
%\begin{center}  
%\begin{equation}  
$r_s=1.4(A_c^{1/3}  + (A_p-1)^{1/3})$    
%\end{equation}  
%\end{center}  
and $a_s=0.6$ fm. For the considered reaction, we have $r_s = 8.1$ fm. 
The maximum impact parameter, corresponding to the impact parameter where the fraction   
of the wave function emitted saturates, is 110   
fm in the case of a Ti target. At this impact parameter the fraction of wave function   
emitted is around 0.04\% and only the noise inherent to the numerical method remains. No   
larger impact parameter is included in the cross section estimate.

\begin{figure}  
\begin{center}  
\includegraphics[width=8cm]{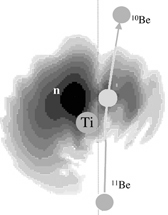}  
\end{center}  
\caption{Probability density of the neutron wave function after the scattering of   
$^{11}$Be onto a Ti target at an impact parameter of 8 fm. The plot is presented in logarithmic   
scale and summed over the coordinate perpendicular to the reaction plan. Also presented   
is the trajectory of the $^{11}$Be and the initial position of the target.}  
\label{dens}  
\end{figure}

The result of the calculation is presented in Fig.\ref{dens} as a probability density plot   
of the neutron for a collision with a Ti target at an impact parameter of 8 fm. We observe   
that a fraction of the wave function is emitted at large angle. This is what is called   
the "towing mode", already  experimentally observed with stable beam \cite{jas98}.  
  
Besides the particle emitted to the continuum, the rest  
of the wave-function either remains in the projectile potential or is transfered to the target.  
Since in that case, the particle is not emitted, these contributions are removed  
to compare with the break-up channel measured in the experiment. In practice, this is performed 
by adding an imaginary   
part to the target and projectile potentials at the end of the calculation.

\section{Experimental results}  
\subsection{Neutron angular distribution in coincidence with $^{10}$Be}  
\label{chap-ds-dW}  
  
\begin{figure}  
\begin{center}  
\includegraphics[width=8cm]{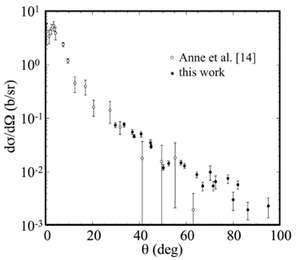}  
\end{center}  
\caption{Angular distributions for neutrons of kinetic energy above 26 MeV. Open dots are   
data from ref.\cite{ann94} and black dots are our data.}  
\label{ds-dW}  
\end{figure}

Fig.\ref{ds-dW} shows the neutron angular distribution in coincidence with $^{10}$Be   
ejectiles and compared to the experimental data of ref.\cite{ann94} with the same neutron energy   
threshold of 26 MeV. The data are corrected for the neutron and the spectrometer detection   
efficiencies. At neutron angles above 30$^o$, due to the use of a spectrometer that allowed   
for the removal of the incident beam before the fragment was detected, much higher statistics   
was gathered allowing for a more precise analysis as will be shown in the following.  
  
\subsection{Angular correlation between the $^{10}$Be and the neutron}  
\label{chap-ds-dW-dte-gau}  
  
\begin{figure}  
\begin{center}  
\includegraphics[width=8cm]{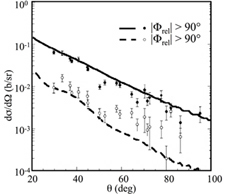}  
\end{center}  
\caption{Neutron angular distribution (after efficiency corrections) for neutron energies above   
26 MeV measured in coincidence with $^{10}$Be ejectiles. Black dots are for events where the neutron   
and the ejectile were measured on opposite sides of the beam ($|\Phi_{rel}| > 90^o$)   
whereas the open circles   
are the events where the neutron and the fragment have been detected on the same side of the beam   
($|\Phi_{rel}| < 90^o$). Corresponding TDSE calculation performed for a $2s$ wave function are represented as   
the plain and the dashed lines, respectively. }  
\label{ds-dW-dte-gau-141106-20-100}  
\end{figure}

A previous analysis \cite{fal02} of the data of ref.\cite{ann94} has shown that the large angle   
neutron emission is mainly due to small impact parameters (b $<$ 15 fm) for which the halo   
wave function is diffracted by the nuclear potential of the target. In such a case, the TDSE   
calculation predicts that the $^{10}$Be and the neutron are preferentially emitted on opposite   
sides of the beam, giving rise to a specific angular correlation as shown in   
Fig.\ref{ds-dW-dte-gau-141106-20-100}. It shows that when the core is   
scattered on the right side of the target, most of the neutron wave function is present   
on the opposite side. In the experiment of ref. \cite{ann94}, the angle of $^{10}$Be were not measured,   
precluding the investigation of such correlations.   
  
In this experiment, such a correlation could be extracted since the scattering angle of   
the ejectile was measured in the SPEG detection system. The experimental neutron angular   
distributions are shown after efficiency correction in Fig.\ref{ds-dW-dte-gau-141106-20-100}   
for events where the ejectile   
and the neutron were detected on opposite sides of the beam axis (black circles,   
$\Phi_{rel}>90^o$) and for events where both the ejectile and the neutron were   
detected on the same side of the beam (empty circles, $\Phi_{rel}<90^o$). We observe an enhancement   
of about a factor 4 for the neutron cross section when detected in coincidence with a $^{10}$Be  
on the opposite side of the beam with respect to the same side of the beam. The solid curve   
results from the calculation  
for a neutron and a core on opposite sides of the initial $^{11}$Be   
direction arbitrarily normalized to the data points (black circles). The dashed line corresponds   
to a core and a neutron flying on the same side. 
Presented calculations were  
performed for a $2s$ wave function but other angular momenta show no difference in neither 
the neutron angular distributions nor the relative asymmetry.  
The relative normalization of these two curves   
is given directly by the calculation. 
In the region between 30$^o$ and 50$^o$, where the cross   
section is large, the calculation reproduces very well the experimental asymmetry. At larger   
angles the actual asymmetry is slightly smaller than the prediction. Overall the anti-correlation   
between $^{10}$Be and neutrons, which is a very detailed feature of the reaction mechanism,   
is remarkably well accounted for.

We thus conclude that the break-up mechanism is well understood and described by our dynamical   
calculation and that the emission of neutrons at large angles is driven by the nuclear interaction   
of the neutron with the target through the so-called "towing mode" mechanism.   
  
\subsection{Gamma spectrum}  
\label{chap-gamma}  
  
\begin{figure}  
\begin{center}  
\includegraphics[width=14cm]{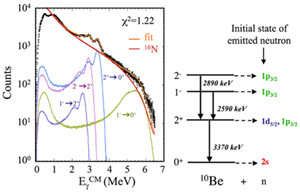}  
\end{center}  
\caption{Left: Gamma energy spectrum in coincidence with a $^{10}$Be and fitted by the contributions of 4   
peaks with the GEANT3 code. Right: Level scheme of $^{10}$Be.}  
\label{gamma}  
\end{figure}

In order to access the final states of the $^{10}$Be core after the break-up, coincidences were measured   
with the gamma rays emitted during the reaction. Figure \ref{gamma} presents the gamma ray spectrum, to which   
a Doppler correction was applied to account for the velocity of the ejectile, recorded in coincidence   
with a $^{10}$Be. It shows three peaks located around 2.6, 2.9 and 3.4 MeV, which correspond to the   
known transitions of $^{10}$Be shown in the level scheme displayed in the right panel. To estimate the   
background, a run with a $^{18}$O beam was performed and the gamma spectrum extracted in coincidence  
with a $^{16}$N detected in the spectrometer. In this reaction, no gamma rays above 1 MeV are expected.   
The whole Ch\^ateau de Cristal configuration was described in the GEANT3 code to infer its response   
to these three observed peaks as well as the direct decay from the 1$^-$ state to the GS which gives   
rise to a gamma ray of 5.96 MeV energy \cite{lim04}. A fit was performed including these four components   
plus the background, with free normalizing factor for each, and reproduces well the experimental   
spectrum above 1.5 MeV.  
  
The level scheme presented in Fig.\ref{gamma} shows that the gamma ray of the decay from the 2$^+$ state to   
the GS is part of the cascade of two other decays from the 1$^-$ and the 2$^-$ states. Therefore, the   
observation of a 3.37 MeV gamma ray signifies that either we started from a 2$^+$ state with a   
neutron hence in a 1d state, or we are measuring the decay from the 1$^-$ or the 2$^-$ states through   
the 2$^+$ state and the emitted neutron stemmed from the $^{10}$Be core, hence from a 1p$_{3/2}$ or a 1s$_{1/2}$   
state. Therefore, the number of 2$^+$ to $0^+$ gamma rays should be equal to the number of cascades   
from the 2$^-$ and the 1$^-$ states plus the number of events feeding directly the 2$^+$ state.

From the fit to the data presented in Fig.\ref{gamma} we extracted that 2/3 of the 3.37 MeV gamma ray   
were the result of a cascade from the 2$^-$ or the 1$^-$ state, whereas 1/3 of the total number of 3.37 gamma   
rays were due to a direct feeding of the 2$^+$ state. This ratio will be used in paragraph \ref{en-3f} when   
analyzing three-fold events.  
  
\subsection{Neutron kinetic energy spectra - 2 fold events}  
\label{en-2f}  
\begin{figure}  
\begin{center}  
\includegraphics[width=8cm]{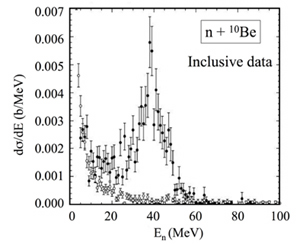}  
\end{center}  
\caption{Neutron energy spectrum in coincidence with a $^{10}$Be and for angles above 30$^o$   
(black dots). Open circles are the data in coincidence with a $^{11}$Be.}  
\label{En-inc}  
\end{figure}

The neutron energy spectrum for two-fold events where a neutron was detected above 30$^o$ in   
the laboratory frame in coincidence with a $^{10}$Be (black dots) is presented in Fig.\ref{En-inc}. It shows a   
peak centered at about 40 MeV and a contribution at low energy.   
  
\begin{figure}  
\begin{center}  
\includegraphics[width=8cm]{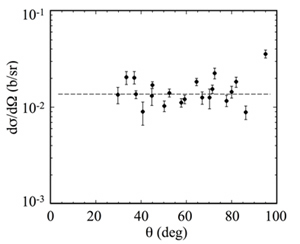}  
\end{center}  
\caption{Angular distribution of neutrons below 15 MeV of kinetic energy.}  
\label{ds-dW-fond}  
\end{figure}

\begin{figure}  
\begin{center}  
\includegraphics[width=8cm]{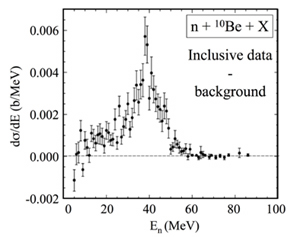}  
\end{center}  
\caption{Neutron energy spectrum in coincidence with a $^{10}$Be and subtracted for the neutrons  
emitted by the target.}  
\label{En-sous}  
\end{figure}

The angular distribution for neutrons above 26 MeV has already been shown in Fig.\ref{ds-dW}   
whereas the angular distribution for neutrons below 15 MeV is presented in Fig.\ref{ds-dW-fond}.   
The latter shows an isotropic behavior in the laboratory frame indicating that these neutrons   
are not coming from the moving projectile but rather from the target at rest. A way of   
estimating the shape of the energy spectrum of the neutrons emitted by the target is to   
extract the neutron spectrum in coincidence with a $^{11}$Be detected in the focal plane. Here,   
apart from a pick-up break-up component which should be very small, one ensures that the   
detected neutron arises from inelastic excitation followed by decay of the target. This   
spectrum is presented in Fig.\ref{En-inc} (open circles) and exhibits only a low energy neutron   
component. We now subtract this component from the spectrum of Fig.\ref{En-inc} (black dots)   
and the result is presented in Fig.\ref{En-sous}. The high energy peak now dominates the spectrum   
with a small tail extending to lower energies.  
   
\subsection{Three-fold events}  
\label{en-3f}  
  
\begin{figure}  
\begin{center}  
\includegraphics[width=14cm]{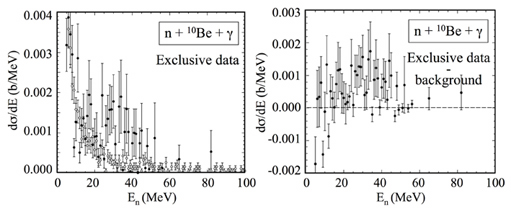}  
\end{center}  
\caption{Neutron energy spectra in coincidence with a $^{10}$Be and a gamma ray of energy between   
1 and 4 MeV. Left: the black dots are the total exclusive data and the open circles are the   
background coming from the target emission and normalized to best fit the black dot spectrum.  
Right: result from the subtraction of the target emission.}  
\label{En-excl-fond-sous}  
\end{figure}

We present on Fig.\ref{En-G} the neutron energy spectrum in coincidence with gamma rays between   
1 and 4 MeV. We estimate from the GEANT3 simulation that the efficiency for detecting the decay of   
the 2$^+$ state is 50\% in this energy range. Furthermore, according to paragraph \ref{chap-gamma},   
1/3 of the 2$^+$ to the $0^+$ gamma-rays come from direct feeding of the 2$^+$ state. Besides the   
low energy component, a broad   
structure is observed, located around 30 MeV. In the same ways as in paragraph \ref{en-2f}, the subtraction   
of the neutron component coming from the target (open circle) is then performed and leads to the right   
panel of Fig.\ref{En-excl-fond-sous} where only the 30 MeV structure remains.   
  
\begin{figure}  
\begin{center}  
\includegraphics[width=8cm]{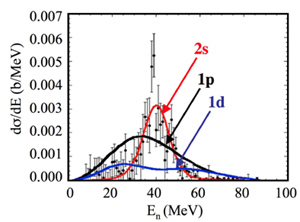}  
\end{center}  
\caption{Neutron energy spectrum when no gamma ray is emitted. Curves are the TDSE calculations   
for a neutron initially in a $2s$, $1p$ and $1d$ state.}  
\label{En-noG}  
\end{figure}

In order now to obtain a neutron energy spectrum for events where no gamma ray was emitted we subtract   
the three-fold spectrum (Fig.\ref{En-excl-fond-sous} right) from the inclusive spectrum of Fig.\ref{En-sous}.   
The result is presented in Fig.\ref{En-noG} as black dots. For this last subtraction, no normalization   
was performed as the detection probabilities were taken into account. A peak clearly shows centered about 40 MeV.   
  
\section{Comparison with the TDSE calculation and possible extraction of spectroscopic factors}  
\label{spefact} 

The TDSE calculation was performed for several initial neutron wave functions   
and results are shown in Fig.\ref{En-noG}. For   
the 1d wave function, the $^{10}$Be potential is no longer spherical. Indeed, the 2$^+$ state of 
$^{10}$Be corresponds   
to a deformed state and was found \cite{aut70}, through inelastic scattering of protons, to have a   
deformation parameter $\beta_2=0.74$. To be able to treat the emission from a deformed core, the   
TDSE model has been adapted by slowly deforming the core potential during the imaginary time   
procedure. The iterations are stopped once the deformation corresponds to the $\beta_2$ value of the   
core and the imaginary time procedure has converged. The depth of the obtained deformed potential   
was adjusted to obtain a binding energy of 3.8 MeV, the sum of the $^{11}$Be binding energy and the 2$^+$   
energy of $^{10}$Be. We performed then the time dependent evolution of the calculation with three   
orientations of the main axis, along X, Y and Z to take into account the random orientation of the   
nucleus when the collision occurs. The sum of the calculations for the 1d state is shown in   
Fig.\ref{En-noG}. TDSE calculations were also performed for the a 1p initial wave functions with a   
binding energy of 6 MeV and the result is shown in Fig.\ref{En-noG}. The calculation also predicts   
that the contribution from a 1s initial wave function is negligible compared to the one of the 1p   
and is not included in the following analysis.

\subsection{Contribution of the $2s$ orbital}  
  
The experimental neutron energy spectrum for which no gamma ray has been emitted is fairly well reproduced by  
the TDSE calculation performed with the $2s$ wave function. 
We clearly see in Fig.\ref{En-noG}, that the TDSE result is highly sensitive to the initial 
angular momentum of the wave-function. It is also worth mentioning that the binding energy of the emitting level  
influences the kinetic energy spectrum. For instance, if the $2s$ wave-function would have been bound by 
$2.75$ MeV, the distribution would have been peaked around $36$ MeV instead of $40$ MeV.

Extraction of spectroscopic factors is performed by comparing experimental absolute cross-sections   
with calculated ones. Here, arbitrary   
normalizations were used for the calculations except for the $2s$ state for which a best fit to   
the data was performed.  
The calculation, assuming that a 100\% of the neutrons are coming from   
the $2s$ state, provides, with the $\chi^2$ minimization method to best reproduce the data, directly the   
spectroscopic factor for this state.  As we have mentionned in section \ref{chap_calc},  
the choice of the minimal impact parameter can influence the result. 
If a sharp cut-off at $b_{min} =8.5$ fm  
is used, the deduced spectroscopic factor is $0.46$ while the prescription of ref. \cite{bon99} 
leading to a smooth 
cut-off gives a
spectroscopic factor of $0.42$.

\subsection{Contribution of the $1p$ and $1d$ orbitals}  
    
\begin{figure}  
\begin{center}  
\includegraphics[width=8cm]{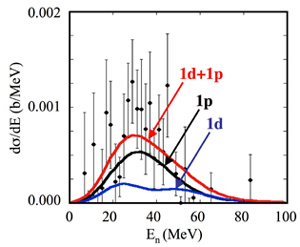}  
\end{center}  
\caption{Neutron energy spectrum in coincidence with a gamma ray of energy between 1 and 4 MeV.   
Curves are the TDSE calculations for a neutron initially in a 1p and 1d state. The sum of the   
two contributions is also presented.}  
\label{En-G}  
\end{figure}

%Keeping in mind the input parameter sensitivity,}  
The same comparison was performed for the three-fold events where a gamma ray is emitted. Experimental   
data are shown in Fig.\ref{En-G}. In that case it is clear that the data can not be reproduced by   
the calculated $2s$ component which peaks at 40 MeV whereas the data are centered at about 30 MeV.   
This precludes the two-step process in which a $2s$ neutron would be emitted together with an excitation
of the core in a $2^+$ state as proposed in ref.\cite{sum06}. 
Both the 1p and 1d contributions present a shape similar to the data. 
However a constraint is that we expect to measure   
twice as many neutrons coming from the 1p state than from the 1d state. Therefore we fixed this ratio between   
the two calculations and normalized the weighted sum to best reproduce the data. This is presented   
in Fig.\ref{En-G} and the respective spectroscopic factors obtained are : $S_{1p}=3.9\pm1.5$ and   
$S_{1d}=0.50\pm0.20$ where the quoted errors are purely statistical.  
%However, it is worth mentioning that the discussion on the  minimal impact parameter to be considered  
%is even more critical for deformed nuclei.  
Here, for the 1d contribution, we have supposed a sharp cut-off between $7$ fm and $9$ fm depending 
on the relative orientation of the  
nuclei.  
%These values are also expected to significantly influence the extracted spectroscopic factor values. 

\subsection{Sensitivity to input parameters on extracted spectroscopic factors for the $2s$ orbital}  
\label{sensitivity}
\begin{figure}  
\begin{center}  
\includegraphics[width=8cm]{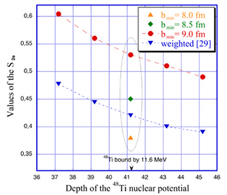}  
\end{center}  
\caption{Variation of the spectroscopic factor with the minimum impact parameter and the absolute value of the
perturbating nuclear potential depth.   
The ellipse indicates values corresponding to the   
$^{48}$Ti potential depth leading to a $1f$ bound neutron at the proper energy.}
\label{S2s}  
\end{figure}

While the shape of the distribution is rather  
insensitive to model parameters,
the absolute cross sections   
appear to be rather sensitive to the inputs of the calculation.  A preliminary study of 
the influence of the parameters   
can be found in ref. \cite{lim04}.  Similarly  to knock-out \cite{han03} or transfer  \cite{lee06} reactions,   
parameters should be carefully optimized to be able to infer spectroscopic factors.   
In Fig.\ref{S2s}, 
we further illustrate the sensitivity of the extracted spectroscopic factor by 
varying the impact parameter   
cut-off and target potential depth.
%However the calculation is quite sensitive to the choice of the minimum impact parameter which   
%consequently changes the value of the spectroscopic factor. It also has some sensitivity to the   
%perturbating nuclear potential that is used.  
%Fig.\ref{S2s} presents the value of the spectroscopic factor when varying either the minimum impact parameter or the  
%depth of the $^{48}$Ti nuclear potential. We remind that a value of -41.2 MeV was chosen in order to obtain a   
%neutron binding energy of 11.6 MeV.   

%If we believe on the sharp cut-off and a minimum value of 8.5 fm for the impact parameter  
%the value extracted here is   
%$S_{2s}=0.45$.   
%When using the "survival factor" presented in ref.\cite{bon99} the spectroscopic factor obtained is   
%even lower ($S_{2s}=0.42$) and goes up to 0.64 for a minimum impact parameter of 9.5 fm.   
%\textcolor{blue}{  
%In addition, for a given impact parameter cut-off, it could be clearly seen in \ref{S2s} that    
%a change in the target potential depth may significantly influence the extracted spectroscopic factor  
%leading to further uncertainty. Altogether, we see that the comparison with the TDSE calculation could only lead   
%to a rather rough estimate of the spectroscopic factor unless parameters are perfectly under control. {ICI IL FAUDRAIT  
%METTRE EVENTUELLEMENT DES BARRES D'ERREUR AUTOUR DE LA VALEUR EXTRAITE!!!}}    

Depending on whether we assume a sharp cut-off in the impact parameter at $b_{min}=8$, $8.5$ and $9$ fm  
or a weighted sum as described in section \ref{chap_calc} the  
spectroscopic factor extracted from the calculation runs from 0.39 to 0.54 for the $2s$ state 
when other parameters remain unchanged. 
We also display in Fig.\ref{S2s} the effect of a change  
in the target potential depth for the sharp cut-off $b_{min}=9$ fm and the weighted case.
The deeper the potential, the lower is the extracted value of the spectroscopic factor.
However, it should be noted that the prescribed value of $41.2$ MeV is optimal to properly  
describe the target binding energy and can not be considered as a free parameter.
  
The sensitivity of the calculation has been further tested with respect to the radius of the 
Wood-Saxon potential of the $^{11}Be$ nucleus.We used $r_0$=1.20 instead of 1.27 and reajusted the depth 
of the potential to obtain a $2s$ state bound by 0.503 MeV. The extracted spectroscopic factor for b$_{min}$=8.5 fm
turned out to be 0.47 instead of 0.46.

Finally, we took the nuclear potential used in ref.\cite{fal02} that has a surface term 
according to N.Vinh Mau's prescription \cite{vin95}. In that case we obtained a lower value, 0.42, probably 
due to the fact that the wave function is on the average positionned more out-side the nucleus core and therefore
a larger fraction 
of it is towed by the nuclear potential.

It is important to note that the present TDSE calculations are made with several  
approximations. First, a simple   
Coulomb trajectory for the target and projectile is supposed. Since we are considering here rather   
small impact parameters, a proper treatment of the nuclear plus Coulomb trajectory is {\it a priori} required.
However, to see the influence of the trajectory on the result, a straight line trajectory was used and the 
extracted 
spectroscopic factor was then 0.48 instead of 0.46. It can then be infered that taking into account a
more realistic trajectory, with the Coulomb and the nuclear potentials, would affect only 
slightly the spectroscopic factors.   
The second simple approximation made regards the fraction of the neutron wave-function initially 
in the projectile   
and transferred to the target. 
This transfer is actually poorly treated in the current TDSE calculation for two   
reasons. First, the effect of the Pauli principle is neglected. Second, a proper treatment of this mechanism   
would also require to add an imaginary potential to the Ti potential.  
In our model, such an imaginary potential is only added {\it a posteriori} to remove the  
fraction of the transferred wave-function when comparing with experimental data.   
%Overall, simplifications in the dynamical treatment add more uncertainty   
%in the extraction of spectroscopic factors.            

\section{Discussion}  
\label{discu}  
\begin{figure} 
\begin{center}  
\includegraphics[width=6cm,angle=-90]{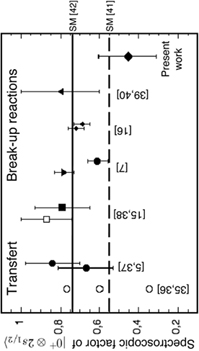}  
\end{center} 
\caption{Spectroscopic factors of the $|0^+\otimes 2s_{1/2}\rangle$ deduced from transfert reactions \cite{zwi79,tim99,win01,for99}
and nuclear and/or coulomb break-up \cite{cha89,aum00,nak94,men97,pal03,fuk04}. The two lines 
indicates the prediction of two different shell model (SM) calculations\cite{ots93,bro01}.} 
\label{palit}  
\end{figure}

The value of the spectroscopic factor $S_{2s}$ obtained in this paper
is plotted in Fig.\ref{palit} (adapted from ref. \cite{pal03}), and compared with  
different experimental data and shell model calculations.
We see that our result is below the average value of other observations
but it should be noted that our error bar is rather large.
During the analysis we had to correct for the small acceptance of the spectrometer 
(see section \ref{chap_setup}). Due to the shape of the cut-off, this correction brings an error which we can
roughly estimate to be between 10 and 20\%. This is based on Fig.\ref{ds-dW} where a comparison is made between the 
present data and a previous measurement. 
A rough estimate of the errors on the spectroscopic 
factor accounting for the acceptance uncertainty, the experimental statistical error and the uncertainty on cut-off in 
impact parameters leads to a spectroscopic factor equal to $0.46\pm 0.15$.   

As discussed in paragraph  \ref{sensitivity}, TDSE calculations performed to extract 
spectroscopic factors are mostly sensitive to  
the cut-off at low impact parameters. 
In order to obtain a spectroscopic factor of 0.8 for the $2s$ state 
we shall use a sharp cut-off of about 10 fm which seems quite high considering the two nuclei.

The values extracted in the transfer reaction by Fortier et al. \cite{for99} (0.84), Windfield et al.\cite{win01} 
with the same data (from 0.67 to 0.8 depending on the optical potential used)
and from break-up reactions at relativistic energies of ref. \cite{pal03} (0.61(5) or 0.77(12) depending on   
whether Coulomb or nuclear break-up is selected), show the 
influence of the experimental method and of the model used to   
describe the collision. Furthermore a value as low as 0.35 has also been predicted for transfer   
reactions \cite{tim99} when including the recoil excitation and the breakup of $^{11}$Be and the variational   
shell model gives a value of 0.55 for the spectroscopic factor of the $2s$ state \cite{ots93}(see dotted line of 
Fig.\ref{palit}).  
  
As for the 1d state, the spectroscopic factor found of 0.50$\pm$0.2 is much higher than the value obtained   
in the transfer reaction of ref.\cite{for99} but again close to the value of 0.40 given by the   
variational shell model \cite{ots93} that insists on the indispensable coupling between the two   
components of the ground state of $^{11}$Be to describe correctly the parity inversion of the   
1/2$^-$ and the 1/2$^+$ states.  
  
\section{Conclusion}  
  
In this paper we have reported on an experiment studying the break-up of $^{11}$Be performed   
at the GANIL facility. The measurement of the $^{10}$Be remnant in coincidence with neutrons emitted   
at large angles ($>30^o$) has allowed to select the nuclear break-up ("towing mode"), and has   
led, with a comparison to a TDSE calculation, 
to the extraction of spectroscopic factors for the ground state of $^{11}$Be.   
  
A value of $S_{2s}=0.46\pm0.15$ is extracted for which   
a large part of the uncertainty comes from the sensitivity to models parameters and experimental errors 
(statistics + systematics). 
We also obtained $S_{1p}=3.9\pm1.5$ and $S_{1d}=0.50\pm0.20$ where the error bars only come from the  
experimental statistics.   
 
Cross sections for the nuclear break-up are of the order of several tens of millibarns in the case of   
weakly bound nuclei such as halo nuclei. If a good knowledge of the minimum impact parameter could be achieved  
%with a good  
%description of the collision  
and as clean as possible experimental data are obtained, we believe that 
this mechanism could be competitive for the study of nuclear spectroscopy of very rare isotopes.   
   
%Furthermore we believe that the information we can extract from such a study is very much complementary   
%to Coulomb break-up and transfer reactions.  
A study of borromean nuclei such as $^6$He and $^{14}$Be 
through nuclear break-up is currently in progress.  
In these cases the measurement of both the neutrons of the halo is performed and compared  
to an appropriate  
description of the two-neutron wave function with   
a TDSE calculation, extended in such a way to include the correlation between the neutrons. Information   
on the two-neutron configuration could then be extracted.   
  
\section{Acknowledgments}  
  
We are grateful to the Swedish collaboration of the NordBall array and colleagues   
from Uppsala for providing their neutron detectors.  
  
% \appendix  
  
% \section{}  
% \label{}  
  
% Bibliographic references with the natbib package:  
% Parenthetical: \citep{Bai92} produces (Bailyn 1992).  
% Textual: \citet{Bai95} produces Bailyn et al. (1995).  
% An affix and part of a reference:  
%   \citep[e.g.][Ch. 2]{Bar76}  
%   produces (e.g. Barnes et al. 1976, Ch. 2).  

\end{document}